\documentstyle[aps,prb,preprint]{revtex}
\begin{document}
\draft
\tighten

\title{ 
Universality and phase diagram around half-filled Landau level
 }
\author{
  L. W. Wong\cite{email} and H. W. Jiang\cite{email2}
  }
\address{
  Department of Physics,
 University of California at Los Angeles, Los Angeles, Ca 90024
  }
\author{
W. J. Schaff
}
\address{
Department of Electrical Engineering, Cornell University, Ithaca, NY 14853
}
\date{\today}

\maketitle

\begin{abstract}
Gated $GaAs/Al_{x}GaAs_{1-x}$ heterostructures were used to determine the low temperature behavior of the two-dimensional electron gas near filling factor $\nu$ = $1\over2$ in the disorder-magnetic field plane.  We identify a line on which $\sigma_{xy}$ is temperature independent, has value $\sigma_{xy}$ = 0.5$\pm$0.02 ($e^{2}/h)$, and a distinct line on which $\rho_{xy}$ = 2$\pm$0.04  $(h/e^{2})$.  The phase boundaries between the Hall insulator and the principal quantum Hall liquids at $\nu$ = 1 and $1\over3$ show levitation of the delocalized states of the first Landau level for electrons and composite fermions.  Finally, data suggests that there is no true metallic phase around $\nu$ = $1\over2$.
\end{abstract}


The study of the quantum Hall systems has been one of the most active areas of research in condensed matter physics during the last decade.  While remarkable progress has been made in understanding the properties of the two-dimensional electron gas (2DEG) in the presence of a magnetic field, the global nature of the quantum Hall systems with regard to the effect of disorder still remains an open question.
	
Theoretically, in the seminal paper\cite{KLZ} of Kivelson, Lee, and Zhang (KLZ), a precise analogy with superfluidity of Chern-Simons bosons was used to derive a set of " law of corresponding states " governing the quantum phase transitions among various quantum Hall liquids (QHL) and (or) Hall insulator (HI).  KLZ constructed a global phase diagram for the 2DEG in the disorder-magnetic field plane, and stressed the issue of universality of phase transitions.  In another elegant theory\cite{HLR} based on the idea of composite fermions in half-filled Landau level, Halperin, Lee, and Read (HLR) also presented a nearly identical phase diagram in the context of discussing the effect of disorder on the composite fermions.
	
Experimentally, the magnetic-field-induced delocalization from HI to a QHL, which is an essential feature of the theoretical global phase diagram, was observed in several $GaAs$ systems\cite{MFI}.  The determination of the phase diagram in the integer quantum Hall effect (IQHE) regime was made in studies of low-mobility gated  $GaAs/Al_{x}GaAs_{1-x}$ heterostructures\cite{GA}, and in studies of Si metal-oxide-semiconductor field-effect transistors (MOSFET's)\cite{SI}.  These studies were extended to the fractional quantum Hall effect (FQHE) regime in an ambitious attempt by Kravchenko \textit{et al}.\cite{KRAVCHENKO} to provide a complete experimental phase diagram. They performed a detailed study on the 2DEG in the Si MOSFET's in the IQHE regime.  Using the ``law of corresponding states'' developed by KLZ, they generalized  their portion of the global phase diagram from the IQHE to the FQHE regime. This approach is, however, debatable. As explicitly pointed out by Kravchenko \textit{et al}., the 2DEG in $GaAs/Al_{x}GaAs_{1-x}$  heterostructures and Si MOSFET's behave very differently in the low field limit.  While a metal to insulator transition is apparently observed in the Si MOSFET's\cite{KRAVCHENKO1}, there is no evidence of a metallic state as B$\rightarrow$0 in  $GaAs/Al_{x}GaAs_{1-x}$ heterostructures\cite{GA}.  There are indeed physical differences in properties among the 2DEG in Si MOSFET's and $GaAs/Al_{x}GaAs_{1-x}$ heterostructures.  For example, the 2DEG in Si MOSFET's has the complication of valley degeneracy, and Zeeman spin-splitting plays a much more important role.  Thus, the one to one correspondence among the properties of the 2DEG in Si MOSFET's and $GaAs/Al_{x}GaAs_{1-x}$ heterostructures as suggested in Ref. 6 needs experimental justification.  Another attempt to determine the phase diagram of the 2DEG in the $GaAs/Al_{x}GaAs_{1-x}$heterostructures was given by Shashkin \textit{et al}.\cite{SHA}.  While the percolation nature of the QHL-HI transition has been revealed in their experiment, the data in the FQHE regime is not conclusive so that the nature of the 2DEG with regard to the disorder in the FQHE regime is still largely undetermined.
	
In this paper, we present an experimental study to probe the topological behavior of the 2DEG in the $GaAs/Al_{x}GaAs_{1-x}$ heterostructures. The $GaAs/Al_{x}GaAs_{1-x}$ system presents perhaps the simplest possible system to determine the phase diagram of the 2DEG, since the band structure of the host lattice has a nearly negligible effect on the properties of the 2DEG.  The behavior of the 2DEG around $\nu$ = $1\over2$ is particularly interesting.  In both phase diagrams proposed by KLZ and by HLR, it is theoretically argued that the topological behavior of the IQHE and the FQHE are identical.  Furthermore, the possibility of the existence of a thermodynamically stable metallic phase at $\nu$ = $1\over2$ in the presence of disorder is postulated\cite{KLZ,HLR}.  These logical predictions are resonable, but certainly require experimental testing.
	
Two sets of samples from two different MBE grown $GaAs/Al_{x}GaAs_{1-x}$ heterostructure wafers were used in our experiments.  The two wafers have undoped $Al_{x}GaAs_{1-x}$ spacer of 50 nm and 80 nm, respectively.  The low-temperature mobility is $\mu\approx 8\times10^{5}cm^{2}$/Vs at a density n $\approx2\times10^{11}/cm^{2}$ for one, and $\mu\approx2\times10^{6}cm^{2}$/Vs at n $\approx2\times10^{11}/cm^{2}$ for the other.  Hall-bar patterns with an aspect ratio of 3.5 : 1 were fabricated by standard photo-lithography, and NiCr gates were evaporated onto the sample surfaces.  The sample densities were controlled by applying a negative bias voltage so that the mobility of the 2DEG can be changed continuously\cite{MFI}.  The magnetotransport measurements were performed by standard low-frequency lock-in techniques in a dilution refrigerator.  The quality of the samples were attested by the large number of the high order fractional states at zero-bias voltage and the high macroscopic homogeneity of the samples.  However, in the interest of determining the disorder effect, the data presented here is concentrated in the regime near the QHL-HI transition where only the principle FQHE states $1\over3$ and $2\over3$ are shown.  Because samples from these two different wafers gave similar results, we will present data from the 50 nm spacer wafer for clarity.
	
The method for constructing the phase diagram in the disorder-magnetic field plane is similar to that used previously for  studies in the IQHE regime\cite{GA}.  To identify the phase boundary between two QHL's, the positions of the dissipative peaks in $\sigma_{xx}$ or $\rho_{xx}$ in the disorder-magnetic field plane are used. Theoretically, it is generally accepted that a dissipative peak in $\sigma_{xx}$ is a signature for the metallic states in the field-theoretical non-linear sigma model\cite{NLSM}. For QHL-HI transition, the phase boundary can be identified either by a temperature-independent (crossing) point in $\rho_{xx}$ or identically by a peak in $\sigma_{xx}$ . The reason is that a peak in $\sigma_{xx}$ should be produced when the Fermi level sweeps across the extended states (note $\sigma_{xx}\rightarrow0$ as $T\rightarrow$0 in QHL and HI phases). 
	
We would first like to present typical raw magnetotransport data for the phase diagram construction.  In Fig. 1, we show examples of traces for $\rho_{xx}$ vs B that we use to determine the HI-QHL phase boundaries. Two classes of transitions are examined : (1) the transition between the $\nu$ = 1 QHL and  HI, and (2) the transition between the primary $\nu$ = $1\over3$ QHL and  HI.  The well-defined crossing points in $\rho_{xx}$ , which were used to construct the phase boundaries, are shown in the figure.  In fact, these two types of transitions were recently studied\cite{SHAHAR}, and the universal values of $\rho_{xx}$ in the critical point were observed\cite{SHAHAR,WJ}.  In this experiment, we found that within our experimental uncertainty, the critical values of both transitions are in good agreement with earlier reports.
	
One of the unexpected findings from this experiment is that we have found two sets of  temperature-independent (crossing) points in both $\rho_{xy}$ and $\sigma_{xy}$ around $\nu$ = $1\over2$.  As shown in Fig. 2, the well defined crossing points are exhibited in both the $\rho_{xy}$ and $\sigma_{xy}$ for a broad range of temperature(from our base temperature 50 mK to about 1.5 K).  In the experiment, the resistivities of both $\rho_{xy}$ and $\rho_{xy}$ were measured directly and the conductivities were obtained through the tensor inversion. In doing the inversion, the conductivity is assumed to be local based on the well-known fact that there is no edge states around $\nu$ = $1\over2$ \cite{DIFF}. Not only are the crossing points well defined, the values of $\sigma_{xy}$ and $\rho_{xy}$  are very close to 0.5 $(e^{2}/h)$ and  2 $(h/e^{2})$, respectively, for a large range of density or disorder (from n = $1\times10^{11}/cm^{2}$ to $4\times10^{10}/cm^{2}$).  Notice that the positions of the crossing points of $\sigma_{xy}$ and $\rho_{xy}$  in B are different in general: the difference becomes large at high disorder.
	
The results of the experimental observations described above are used to form the phase diagram in  Fig. 3.  Following conventions in the literature, the vertical axis is plotted with a variable of {\it $n_{i}$/n} which describes qualitatively the degree of disorder.  The horizontal axis is normalized by the inverse filling factors of {\it eB/nhc}.  The density in this case is determined by the combination of low-field Hall resistance and Shubnikov-de Haas oscillations. 
	
Now, we would like to discuss the underlying physics of this phase diagram.  Some of the features can be interpreted by means of existing models.  Let us start with the phase boundary in the left portion of the diagram.  The solid squares in the Fig. 3 represents the transition points between $\nu$ = 1 QHL to HI. Theoretically,  according to the scaling theory\cite{NLSM} the values of the conductance of both $\sigma_{xx}$ and $\sigma_{xy}$ should be 1/2 ($e^{2}/h$). We found the actual values of $\sigma_{xx}$ and $\sigma_{xy}$ at the phase boundary to be $0.5\pm0.1   (e^{2}/h)$.  It is also well known that the lowest delocalized state in the low-disorder limit follows the lowest Landau level at $\nu$ = $1\over2$.  The observation that the phase boundary of $\nu$ = 1 QHL deviates from $\nu$ = $1\over2$ is a signature of the levitation of the lowest delocalized state in high disorder\cite{GA}.  These solid squares merge continuously with the crossing points in $\sigma_{xy}$, represented by the open circles, as disorder decreases. The existence of universal value in $\sigma_{xy}$ in the weak disorder (noninsulating) regime is however not expected.  A recent theory, considering the electron-hole symmetric disorder, predicted that $\sigma_{xy}$ is universal\cite{KIVELSON} at precisely $\nu$ = $1\over2$ in the lowest Landau level.
	
Next, we would like to discuss the significance of crossing points with $\rho_{xy}$ = 2 $(h/e^{2})$ represented by the open squares in Fig. 3.  We believe it is a result predicted by the composite-fermion theory.  In the composite-fermion formalism\cite{HLR,KAL}, the physical resistivities are given by $\rho_{xx}$ = $\rho^{cf}_{xx}$(T)   and $\rho_{xy}$ = [2 + $\rho^{cf}_{xy}$(T)] in the units of $h/e^{2}$.  Within this frame work, $\rho_{xy}$(T) = 2 $(h/e^{2})$ at $\nu$ = $1\over2$ implies that the $\rho^{cf}_{xy}$(T) = 0 (i.e., the composite fermions experience a zero effective field). The seemingly puzzling feature of the data in Fig. 3 is that the $\rho_{xy}$ crossing-point line deviates from the $\nu$ = $1\over2$ line (i.e., the exact ratio of two flux to one electron) as disorder is increased.  
        
We can offer a semiquantitative explanation for this behavior.  As mentioned, the position of $\nu$ = $1\over2$ in Fig. 3 is determined by  using the density obtained from the low-field measurements.  However, in the presence of strong mixing of the spin-up and down Landau levels, the actual position of $\nu$ = $1\over2$ could be substantially different from those in Fig. 3.  As illustrated graphically by Fig. 4 in the weak mixing limit, almost all the electrons in the lowest Landau level are in the spin-up state.  With stronger disorder, the contribution to the electron population from the spin-down state increases significantly.  Hence, the actual position where the ratio of two flux to one electron is exact happens at lower energy $\varepsilon_{\nu=1/2}$, which is given by 1/2(B/$\phi_{0}$) = \(\int_{-\infty}^{\varepsilon_{\nu=1/2}}\rho(\varepsilon)d\varepsilon\) ,  where $\phi_{0}$ = $hc/e$ and $\rho$($\varepsilon$) is the density of states.  The lower $\varepsilon_{\nu=1/2}$  corresponds to an apparently higher B value in the measurement; thus, the actual position of $\nu$ = $1\over2$ shifts from the $\nu$ = $1\over2$ line in Fig. 3.  In fact we can estimate the level broadening of a Lorentzian shape DOS by using the high-temperature scattering time measured around $\nu$ = $1\over2$\cite{DOS}.  By assuming the spin-splitting of the lowest Landau level is given by the exchange interaction\cite{USHER}, the mixing would produce shift of about 15\% in B at n $\approx5\times10^{10}/cm^{2}$ which is consistent with discrepancy shown in the Fig. 3.  Based on this analysis, we believe the temperature-independent $\rho_{xy}$ = 2 $(h/e^{2})$ line presents a line of composite fermions at zero magnetic field.
	
Let us now turn our attentions to the right portion of the phase diagram, the $\nu$ = $1\over3$ to HI transition.  Like the case of  $\nu$ = 1 QHL to HI transition, the substantial deviation of the phase boundaries from $\nu$ = $1\over4$, can be viewed as a departure of the delocalized state of the composite fermions from its first Landau level.  It is important to note that although the deviation is substantial, no levitation above the Fermi level is observed (i.e., the phase boundary curve downward).  As a consequence, no HI phase on the low-field side of $\nu$ = $1\over3$ QHL is seen.  We attribute this to the finite size or finite temperature of the sample (i.e., the appearance of the noninsulating regime). Much like the delocalized states in the high Landau levels of the IQHE regime\cite{GA}, the observation of the levitation requires not only that the localization length  be smaller than the effective sample size, but also that the size of the gap for the QHL must be larger than the disorder potential.  These requirements are experimentally hard to match in the FQHE regime due to the small size of the energy gap for the FQHE states.
	
Thus far, we have shown the similar global behavior of the first Landau level for electrons and composite fermions in disorder. As mentioned early, the determination of whether there is a thermodynamically stable metallic state around $\nu$ = $1\over2$ is of general importance to the phase diagram.  In order to understand the behavior of the 2DEG around $\nu$ = $1\over2$, we have performed a series of temperature-dependent measurements with the resistivity at $\nu$ = $1\over2$ at different degrees of disorder.  In Fig. 5, several $\rho_{xx}$ curves are plotted as a function of temperature at different densities n.  In the weak disorder regime where the crossing points in $\sigma_{xy}$ and $\rho_{xy}$ are observed, $\rho_{xx}$  can be well described by the form $\rho_{xx}\sim$ $ln(T)$(see insert a). As pointed by HLR, this $ln(T)$ dependence of $\rho_{xx}$ is a consequence of the interactions between the composite fermions in the presence of disorder having long-range Coulomb interactions (also see Rokhinson \textit{et al}.\cite{ROK}). As disorder is increased so that open circles meet with the solid squares, $\rho_{xx}$ exhibits exponential behavior at the low temperature end, and logarithmic behavior at the high temperature end. Finally, $\rho_{xx}\sim$ $exp(1/T)^{1/2}$ in the strong disorder regime(see insert b).  Evidently, this general form for the variable-range hopping in high B\cite{POL} is found in our experiment. The dashed line in Fig. 3 represents an ill-defined crossover from the weak disorder(noninsulating) to the strong disorder regime(insulating).  Therefore, following the development of $\rho_{xx}$ as a function of temperature, we believe that disorder slowly drives the 2DEG towards HI. Thus, it implies that in the limit  of $T\rightarrow0$, there is no true metallic phase around $\nu$ = $1\over2$ at least in the range of disorder where our experiment was performed.
	
In summary, we have experimentally mapped out the global phase diagram around $\nu$ = $1\over2$ in a broad range of disorder.  The observation of the temperature-independent (crossing) points $\rho_{xy}$ = $2\pm0.04$ $(h/e^{2})$ presents evidence that the composite fermions experience zero effective field at $\nu$ = $1\over2$.  Our data shows that there is no true metallic phase around $\nu$ = $1\over2$ at least in the range of disorder where our experiment was performed.  Finally,  the $\nu$ = $1\over3$ QHL to HI phase boundary shows a weak levitation of the delocalized states of the composite fermions, as similar to the $\nu$ = 1 QHL to HI phase boundary.
	
We would like to thank S. Kivelson and B. Shklovskii for useful discussions; and D. Chang for sample preparation.  This work is supported by the NSF under grant number DMR-93-13786.

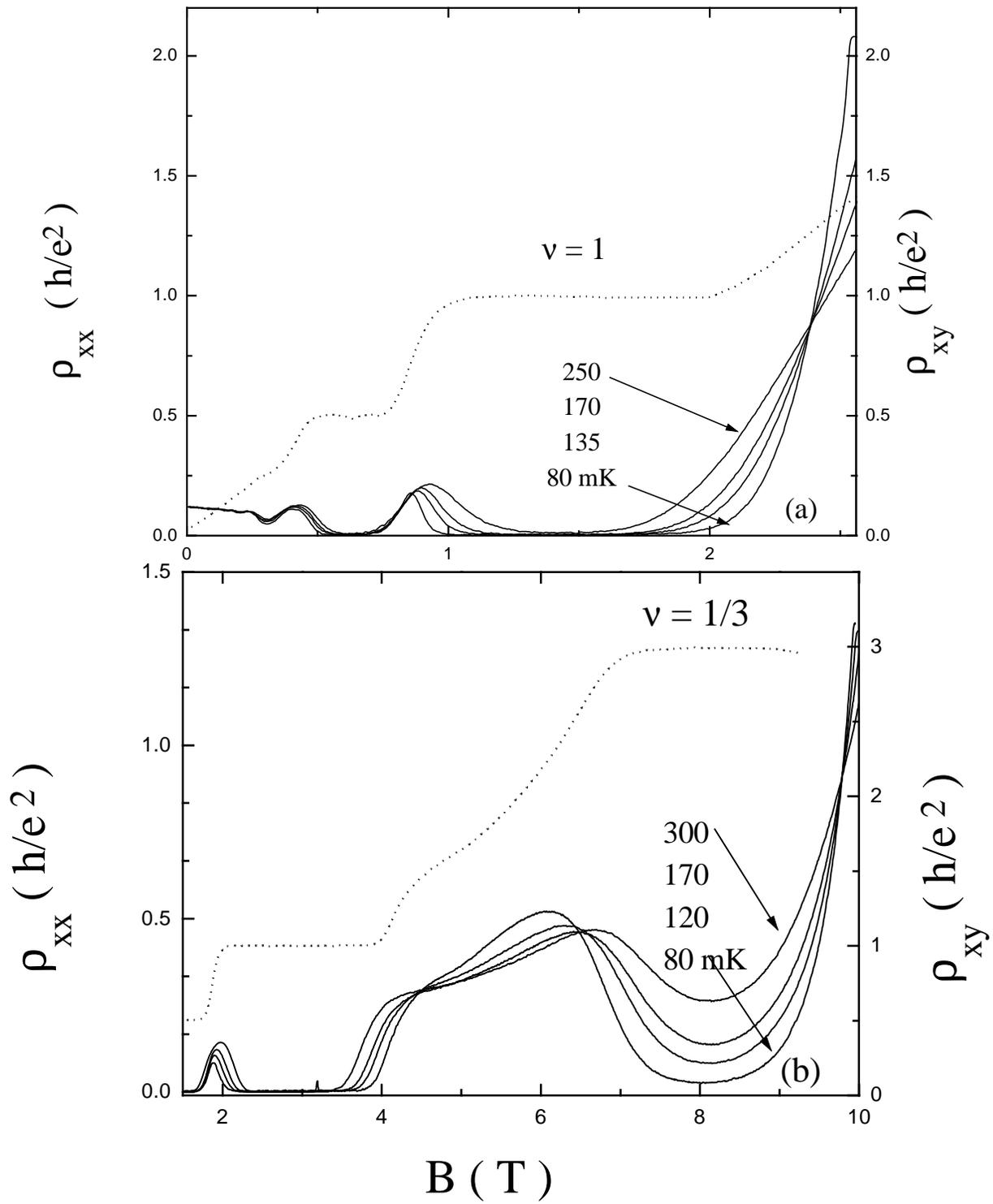
\begin{figure}
\caption{
Typical magnetoresistivities vs magnetic field data at different temperatures for the determination of phase boundary for (a) $\nu$ = 1 QHL-HI transition, and (b) $\nu$ = $1\over3$ QHL-HI transition.  Notice that the temperature-independent (crossing) points are very well defined.
}
\end{figure}
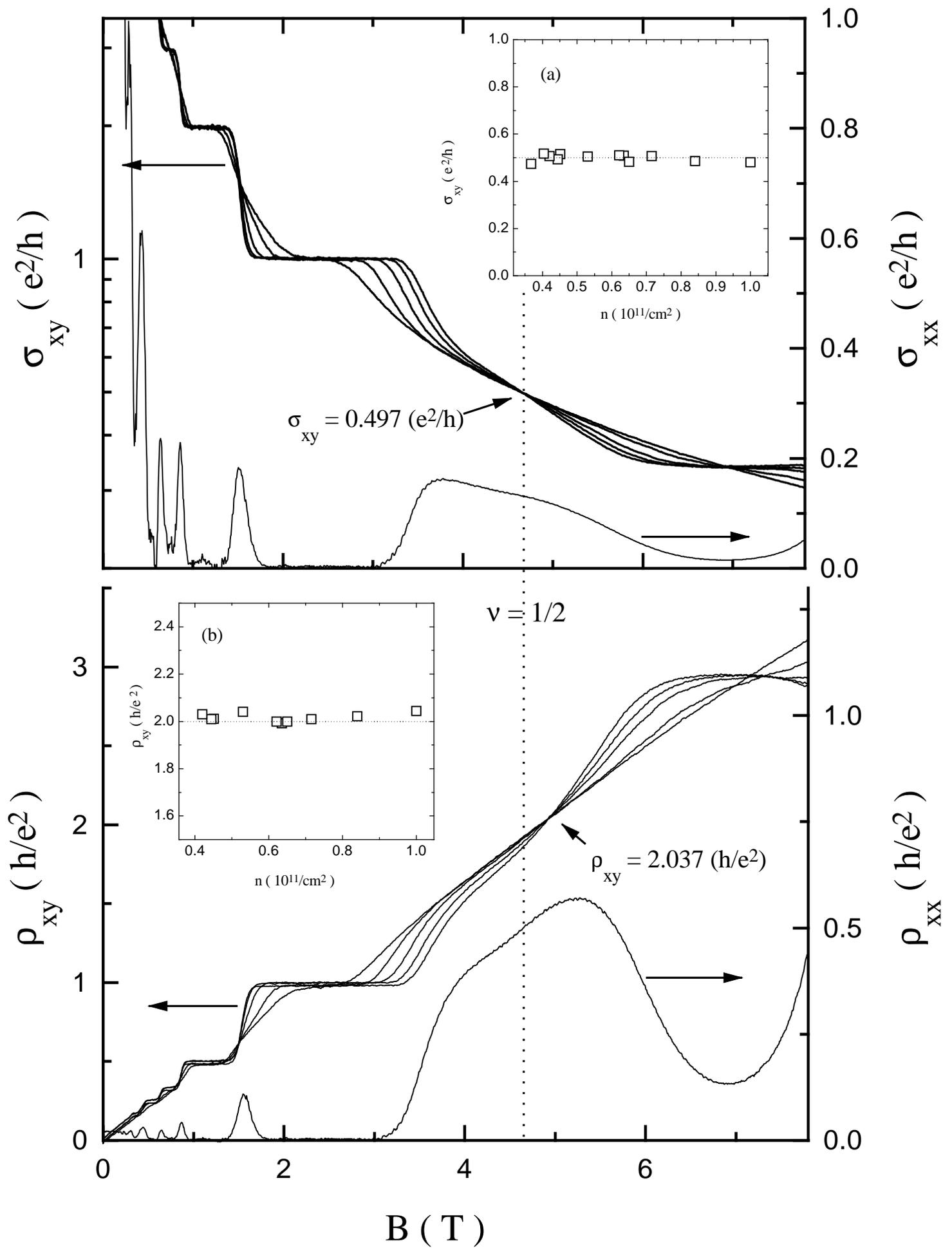
\begin{figure}
\caption{
Traces of the magnetotransport coefficients showing the crossing points near $\nu$ = $1\over2$ at five different temperatures: 1200, 600, 250, 130, 80 mK. Inserts (a) and (b)  show the values of $\sigma_{xy}$ and $\rho_{xy}$ , respectively, at the crossing points for different densities.
}
\end{figure}
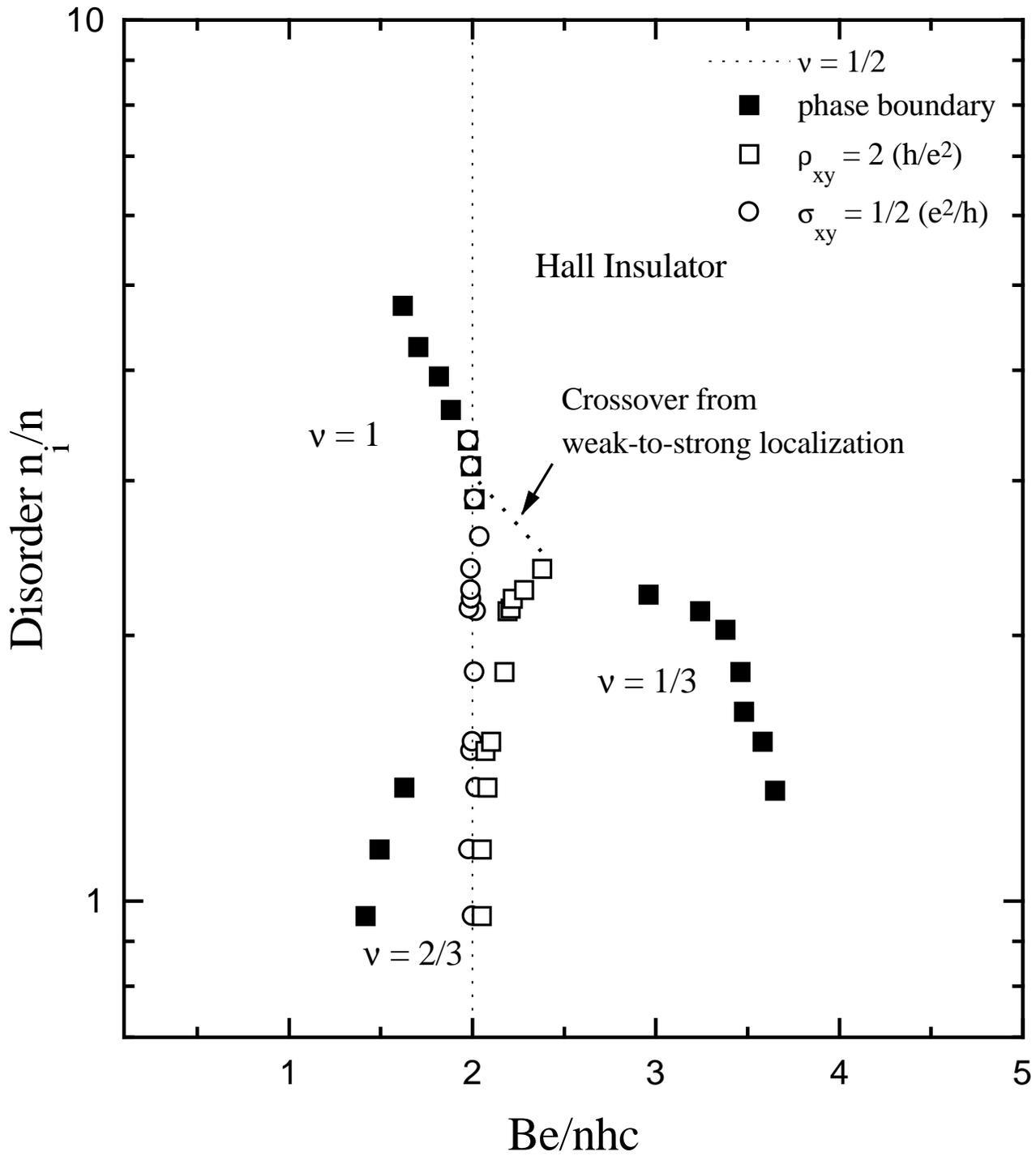
\begin{figure}
\caption{
Experimental phase diagram around $\nu$ = $1\over2$ of the 2DEG in the disorder-magnetic field plane.  The vertical axis, {\it $n_{i}$/n}, where \textit{$n_{i}$} = $1\times10^{11}/cm^{2}$, is meant to represent qualitatively the amount of disorder.  The horizontal axis representing the strength of the magnetic field is normalized to 1/$\nu$.
}
\end{figure}  
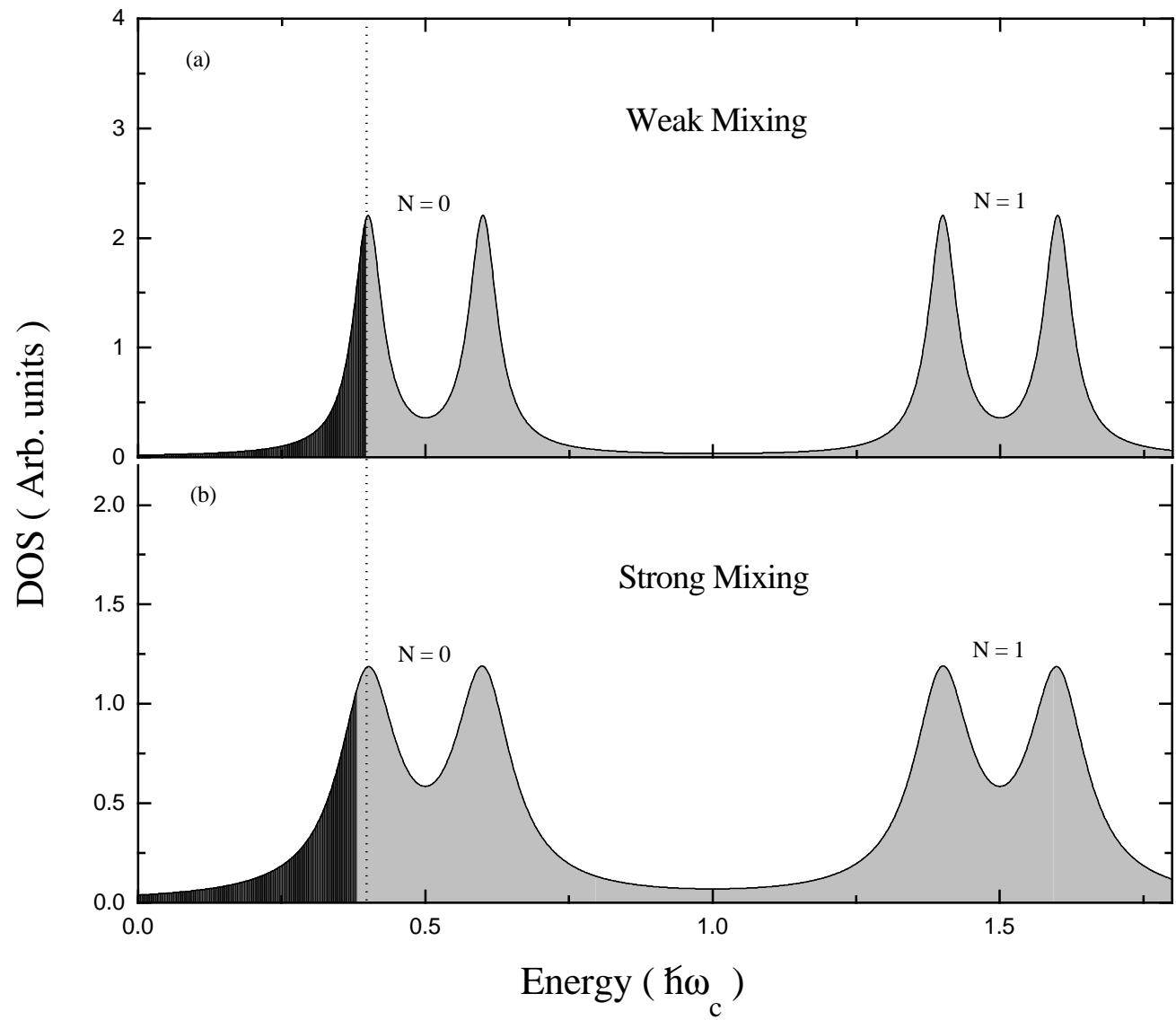
\begin{figure}
\caption{
Graphical illustration of the position of $\varepsilon_{\nu=1/2}$, where the ratio of two flux to one electron is exact, shifts towards lower energy due to the mixing of the spin-up and down disorder-broadened Landau levels.
}
\end{figure}
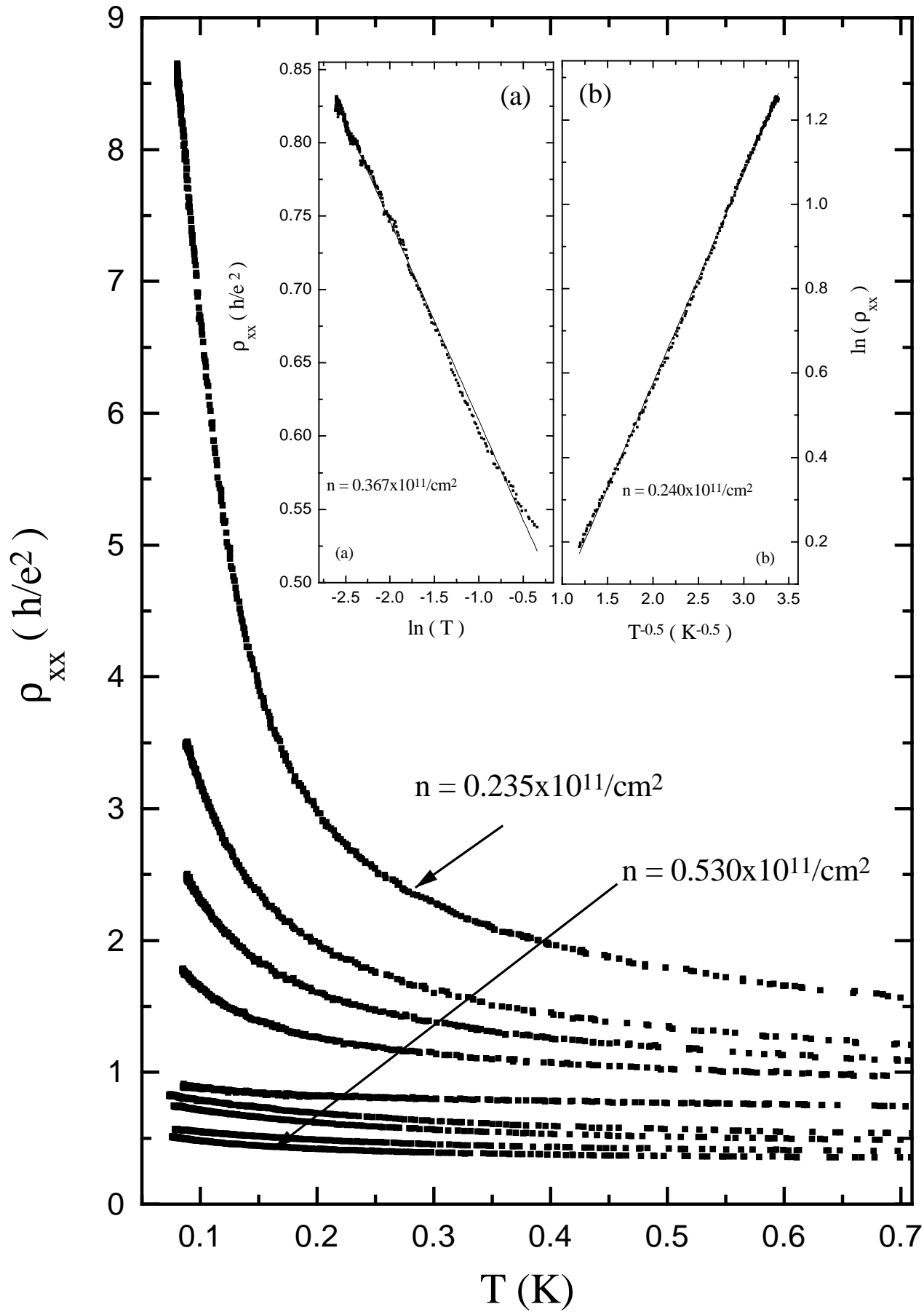
\begin{figure}
\caption{
Temperature dependence of $\rho_{xx}$ at $\nu$ = $1\over2$ with different densities: n = 0.235, 0.240, 0.247, 0.256, 0.304, 0.367, 0.401, 0.444, $0.530\times10^{11}/cm^{2}$. Insert (a) shows the logarithmic dependence of $\rho_{xx}$ in T in weak disorder regime. Insert (b) shows the exponential dependence of $\rho_{xx}$ in T in strong disorder regime.
}
\end{figure}
\end{document}